\documentclass[traditabstract]{aa}
\usepackage{graphicx}
\usepackage{natbib}
\usepackage{txfonts}
\bibpunct{(}{)}{;}{a}{}{,}
\begin{document}

  \title{On the red giant branch mass loss in 47~Tucanae: Constraints from the horizontal branch morphology}
  
  \author{Maurizio Salaris\inst{1}, Santi Cassisi\inst{2,3} \and Adriano Pietrinferni\inst{2}}

\institute{Astrophysics Research Institute, 
           Liverpool John Moores University, 
           IC2, Liverpool Science Park, 
           146 Brownlow Hill, 
           Liverpool L3 5RF, UK, \email{M.Salaris@ljmu.ac.uk} 
            \and
           INAF~$-$~Osservatorio Astronomico di Teramo, Via M. Maggini, I$-$64100 Teramo, Italy, 
            \email{cassisi,pietrinferni@oa-teramo.inaf.it}  
            \and
            Instituto de Astrof{\'i}sica de Canarias, Calle Via Lactea s/n, E-38205 La Laguna, Tenerife, Spain
           }

 \abstract{We obtain stringent constraints on the actual efficiency of mass loss for red giant branch stars in the Galactic globular
 cluster 47~Tuc, by comparing synthetic modeling based on stellar evolution tracks with the observed distribution of stars along the 
horizontal branch in the colour-magnitude-diagram. We confirm that the 
observed, wedge-shaped distribution of the horizontal branch can be reproduced only by accounting for a range of initial He abundances 
--in agreement with inferences from the analysis of the main sequence-- and a red giant branch mass loss with a small dispersion. 
We have carefully investigated several possible sources of uncertainty 
that could affect the results of the horizontal branch modeling, stemming from uncertainties in both stellar model computations and the 
cluster properties such as heavy element abundances, reddening and age.
We determine a firm lower limit of $\sim$0.17$M_{\odot}$ for the mass lost by red giant branch stars, corresponding to horizontal branch 
stellar masses between $\sim$0.65$M_{\odot}$ and $\sim$0.73$M_{\odot}$ (the range driven by the range of initial helium abundances). 
We also derive that in this cluster the amount of mass lost along the asymptotic giant branch stars is comparable to the 
mass lost during the previous red giant branch phase.
These results confirm for this cluster the disagreement between colour-magnitude-diagram analyses and 
inferences from recent studies of the dynamics of the cluster stars, that 
predict a much less efficient red giant branch mass loss. A comparison between the results from these two techniques 
applied to other clusters is required, to gain more insights about the origin of this disagreement.
}
\authorrunning{M. Salaris et al.}
\titlerunning{47~Tucanae mass loss and horizontal branch morphology}
  \maketitle


\section{Introduction}

Mass loss during the red giant branch (RGB) evolution of globular cluster (GC) stars has a generally negligible  
effect on their structure \citep[unless the RGB star is experiencing very high mass loss rates, see][]{cas}, 
but it is crucial to interpret the colour-magnitude-diagram (CMD) of 
the following horizontal branch (HB) phase, and affects the CMD and duration of the 
asymptotic giant branch (AGB) stage.

A comprehensive physical description of RGB mass loss processes is still lacking, and 
RGB mass loss rates are customarily parametrized 
in stellar evolution calculations by means of simple relations like the Reimers formula \citep{reimers}, or more recently 
the \citet{sc05} one. These prescriptions are essentially scaling relations between mass loss rates and 
global stellar parameters like surface bolometric luminosity ($L$) and gravity ($g$), effective temperature ($T_{eff}$) and/or 
radius ($R$). The zero point of these scaling relations is typically set by a free parameter ($\eta$) that needs to be calibrated. 

The more direct approach to study the mass loss in RGB stars is to detect outflow motions in the outer regions of the 
atmospheres \citep[for example the presence of asymmetries and coreshifts in chromospheric lines, see, e.g.,][]{mcp06, vmc11}, 
or detect the circumstellar envelopes at larger distances from the stars 
\citep[for example through infrared dust emission, see, e.g.,][]{ori07, ori10, boyer10, momany}.
Another traditional indicator of the efficiency of RGB mass loss is the CMD location and morphology of the HB of globular clusters, 
starting 
from the pioneering works by \citet{ir70} and \citet{r73}. Matching observed HBs with synthetic HB models traditionally requires 
that RGB stars lose a fraction of their initial mass, with typical 
values of the order of $\sim$0.2 $M_{\odot}$
\citep[see, e.g.,][and references therein]{ldz90, catelan, s07, dicri, gcb10, dale, mcdon}.

A very recent series of papers \citep[][]{heyl_a, heyl_b} has applied a completely different approach to estimate 
the mass lost by RGB stars in the Galactic GC 47~Tuc. Using $HST$ images these authors determined the 
rate of diffusion of stars through the cluster core, using a sample of bright white dwarfs (WDs).
They then compared the radial distribution of upper main sequence (MS), RGB and HB stars, showing that they are nearly identical, 
even when only objects near the RGB tip are considered,  
whilst the radial distribution of young WDs is only slightly less concentrated than upper MS and RGB stars, 
indicating that there has been very little time for 
the young WDs to have diffused through the cluster since their progenitors lost mass. They estimated that most of 
the $\sim$0.4~$M_{\odot}$ that 47~Tuc stars lose between the end of the MS 
(the typical MS turn-off mass for 47~Tuc as inferred from theoretical isochrones is equal to  
$\sim$0.9$M_{\odot}$) and the beginning of the WD sequence  
\citep[typical masses for bright WDs in GCs as determined from observations are $\sim$0.53$M_{\odot}$, see][]{kalwd} is shed shortly before the start 
of the WD cooling.
Quantitatively they estimated that mass loss greater than 0.2$M_{\odot}$ earlier than 20~Myr before the termination 
of the AGB can be excluded with 90\% confidence, and 
that mass loss larger than 0.2$M_{\odot}$ during the RGB can be excluded at more than 4$\sigma$ level. Also, a typical HB stellar mass of 
the order of $\sim$0.65$M_{\odot}$ is excluded by comparisons of the radial distribution of HB stars and $\sim$0.65$M_{\odot}$ MS stars.

Regarding more direct estimates of RGB mass loss in 47~Tuc, 
\citet{ori07} derived that the total mass lost by individual RGB stars is  
$\Delta M_{\rm RGB}\sim 0.23\pm0.07 M_{\odot}$, from the detection of their circumstellar envelopes by means of mid-IR photometry 
\citep[see also][]{ori07, boyer10, momany}. This is in contrast with \citet{heyl_b} result, but is 
consistent with published results from synthetic HB modeling, that require RGB stars to have lost  
typically more than 0.2$M_{\odot}$ \citep[][]{s07, dicri, gratton}. 
On the other hand, observations of the infrared excess around nearby RGB stars (not in GCs) led \citet{groen} to determine 
a Reimers-like mass mass loss formula that when used in stellar model calculations predicts negligible mass loss for 47~Tuc RGB stars 
\citep{heyl_b}.

\citet{heyl_b} result based on stellar dynamics clearly questions the accuracy of HB stellar models and/or their interpretation of HB morphologies. 
A solution of this discrepancy requires a robust assessment of the reliability of these two radically different techniques employed to determine 
the cluster's RGB mass loss.  
To this purpose we revisit in this paper the theoretical modeling of 47~Tuc HB, discussing various sources of potential uncertainties 
in synthetic models based on HB evolutionary tracks.
Section~2 describes briefly our synthetic modeling and presents our {\sl baseline} synthetic HB for this cluster, with the estimated mean RGB mass loss 
($\Delta M_{\rm RGB}$). 
Section~3 discusses various potential sources of uncertainties in our baseline synthetic HB model, and their impact on the estimated $\Delta M_{\rm RGB}$. 
A critical discussion and conclusions close the paper.

\section{The baseline synthetic HB model for 47~Tuc}
\label{refsim}

We have employed in our analysis the accurate $BVI$ cluster photometry by \citet{bs}, and selected 
stars between 400 and 900 arcseconds of the cluster centre, to minimize the effect 
of blending \citep[see the discussion in][]{bs} and field contamination. We present results for the HB modeling in  
the Johnson $V-(B-V)$ CMD, but we have verified that we reach the same conclusions when using the 
Johnson-Cousins $V-(V-I)$ CMD instead.
We have also compared the $V-(V-I)$ CMD of the cluster central 
region\citep[transformed from the equivalent $HST$ ACS filters to the Johnson-Cousins system by][]{acs} 
from the ACS survey of Galactic GCs with our adopted photometry. The HB morphology is the same in the ACS field, with just an offset 
of about $\sim$0.05~mag in the $V$ magnitudes (ACS magnitudes being brighter). 

The reddening estimates for 47~Tuc range between $E(B-V)$=0.024 \citep{gbc03} and $E(B-V)$=0.055 \citep{gratt97}.
\citet{sfd} reddening maps provide $E(B-V)$=0.032, whilst \citet{harris} catalogue of GC parameters reports E(B-V)=0.04. 
The amount of differential reddening is negligible, $E(B-V)$ varies around the cluster mean value by at most $-$0.007~mag and +0.009~mag respectively, 
as recently determined by \citet{mmc} on a sample of stars taken from our adopted \citet{bs} photometry.
As for the cluster chemical composition, \citet{thompson} and \citet{cpj} list a 
series of spectroscopic determinations of [Fe/H] and [$\alpha$/Fe] for this cluster that 
can be summarized as [Fe/H]=$-0.7\pm0.1$ and [$\alpha$/Fe]=0.3--0.4.
[Fe/H] measurements in a sample of 47~Tuc HB stars provide [Fe/H]=$-0.76\pm0.01$ (rms = 0.06 dex), consistent with 
the range quoted above \citep{gratton}.

In our baseline simulation of the cluster HB we have assumed [Fe/H]=$-$0.70, [$\alpha$/Fe]=0.4 and $E(B-V)$=0.024.
We employed the lowest estimate of the cluster reddening because it leads to 
generally higher HB masses, hence minimizes 
the necessary RGB mass loss\footnote{Higher reddenings 
require a larger shift to the red of the HB models, hence a lower HB mass at a given colour, and 
an increased amount of RGB mass loss to match the observed HB location}.

Synthetic HB models have been computed by employing HB tracks from the 
BaSTI stellar model library \citep[][]{basti, bastia}\footnote{http://www.oa-teramo.inaf.it/BASTI}, and 
the code fully described in \citet{dale}. We made use of the BaSTI 
tracks for [Fe/H]=$-$0.7, [$\alpha$/Fe]=0.4 (corresponding to $Z$=0.008), and varying $Y$.
 
Our calculations require the specification of four parameters, plus the cluster initial composition, age, and the photometric error. 
Two of these parameters are related to the distribution of the initial He abundances among the cluster stars. We 
can choose between a Gaussian distribution with a mean value $<Y>$ and spread $\sigma(Y)$, and a uniform distribution 
with  minimum value $Y_{\rm min}$ and range $\Delta Y$.
The other two parameters are the mean value of the mass lost along the RGB, $\Delta M_{\rm RGB}$ --that for simplicity 
we assume to be the same for each $Y$, but can be made $Y$-dependent-- and the spread around this mean value ($\sigma(\Delta M_{\rm RGB})$). 
The idea behind this 
type of simulations \citep[see, e.g.,][and references therein]{dcm, gcb10, dale, m14} 
is that the colour extension of the HB is driven mainly by the variation of $Y$ rather than mass-loss efficiency. 
A range of He-abundances within individual clusters is expected theoretically given the well-established presence 
of CN, ONa, MgAl abundance anticorrelations within single GCs \citep[see, i.e.,][]{gsc}. 
These abundance variations are most likely produced by high-temperature CNO cycling, hence one also expects 
He variations in addition to these anticorrelations.The actual amount of He variations depend on 
the nucleosynthetic site and the cluster chemical evolution \citep[see, e.g.,][for two different 
scenarios to explain the observed abundance patterns]{decress, dercole}.

Indeed, studies of the optical CMDs of MS stars\footnote{Light element anticorrelations 
do not affect the bolometric corrections for MS stars 
in optical CMDs, as shown by \citet{sbordone}} have disclosed the presence of ranges of initial He  
in several GCs \citep[see, e.g.,][and references therein]{p07, m13, nardiello}, including 47~Tuc \citep{milonetuc}.
The synthetic HB modeling by \citet{dicri} and \citet{gratton} also required a range of initial $Y$ to reproduce the wedge-shaped 
HB in optical filters.
On this issue, it may be worth recalling that in the past \citep[see][]{dvl, cfp} it was shown that 
at metallicities typical of 47~Tuc a wedge-shaped HB could be reproduced theoretically with a single but large initial $Y$, of the order of 
$Y\sim$0.30. With our synthetic HB calculations we can obtain a shape roughly similar to the one observed for $Y$=0.34 
and a negligible $Y$ range (and $\Delta M_{\rm RGB}\sim$0.12$M_{\odot}$). However, the resulting distance modulus is  
$\sim$0.3~mag too large compared to constraints from the cluster eclipsing binaries (see below), and in addition cluster  
R-parameter studies \citep[see, e.g.,][]{csi, srcp} exclude such high initial values of $Y$.

Finally, it is also important to mention that the observed CNONaMgAl abundance variations do not affect the stellar evolution tracks and 
isochrones 
as long as the CNO sum is unchanged \citep[see, e.g.,][]{basticno:09,cassisi:13} 
as generally true, within the spectroscopic measurements errors, with just a few exceptions. 
This justifies the use of standard $\alpha$-enhanced models, with just varying initial He content.

To translate the RGB mass loss $\Delta M_{\rm RGB}$ into HB masses we have to assume an age for the cluster, that provides 
(from the theoretical isochrones) the initial value of the mass of the stars evolving at the tip of the RGB (denoted as RGB progenitor mass).
Age estimates by \citet{sw02}, \citet{gbc03}, \citet{dotter10}, \citet{vdb13} from CMD analyses 
provide a range between t=$\sim$10.5 and t=$\sim12.5$~Gyr, 
and we assumed for this simulation t=11.5~Gyr, that corresponds to a RGB progenitor mass equal to 0.92$M_{\odot}$.
Our assumed age is also consistent with the estimated value 11.25$\pm$0.21(random)$\pm$0.85(systematic) Gyr  
by \citet{thompson}, based on theoretical mass-radius relations applied to the cluster eclipsing binary V69. 

\begin{figure}
\centering
\includegraphics[width=\columnwidth]{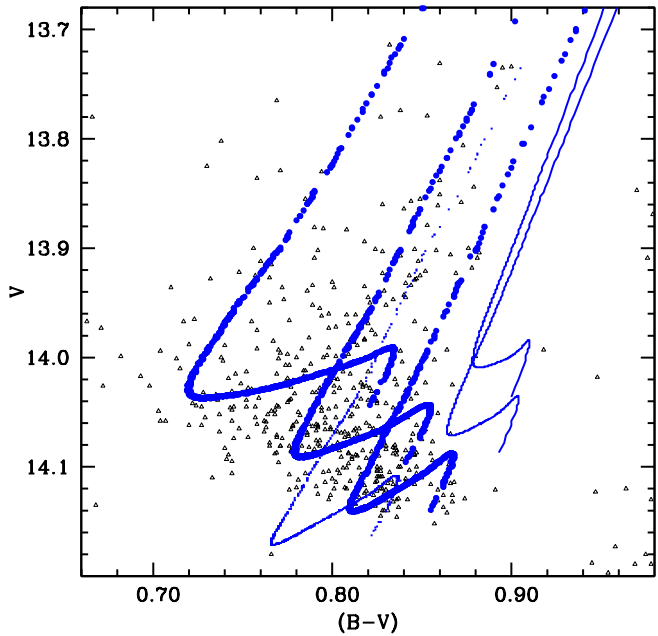}
\caption{Comparison between the observed HB of 47~Tuc (triangles) with synthetic CMDs calculated with $\Delta M_{\rm RGB}$=0.23$M_{\odot}$  
and $Y$=0.256, 0.270 and 0.286 (filled large circles) respectively, plus 
a simulation for $\Delta M_{\rm RGB}$=0.28$M_{\odot}$ and $Y$=0.256 (dots). Two HB tracks corresponding to and $Y$=0.256, and masses M=0.8 and 
0.9$M_{\odot}$  (corresponding to $\Delta M_{\rm RGB}$=0.12$M_{\odot}$ and 0.02$M_{\odot}$, respectively)  
are displayed as solid lines. All the tracks and synthetic CMDs are shifted by $E(B-V)$=0.024 and $(m-M)_V$=13.40 (see text for details).}
\label{HBtracks}
\end{figure}

Figure~\ref{HBtracks} displays a first test that shows clearly the need for a substantial mass loss and a range of initial $Y$ values  
to match the location and morphology of the cluster HB.
The three narrow synthetic sequences in the figure (filled circles) that overlap with the observed CMD have been calculated 
for $Y$=0.256 (the {\sl normal} initial He for the chosen initial metallicity according to the ${\rm \Delta{Y}/\Delta{Z}\sim1.4}$ ratio employed in the 
BaSTI calculations, and a cosmological $Y$=0.245), 0.270, 0.286 respectively, $\Delta M_{\rm RGB}$=0.23 and a negligible Gaussian spread $\sigma(\Delta M_{\rm RGB})$=0.001. 
The synthetic sequences have been shifted in colour by applying the reference reddening $E(B-V)$=0.024, and 
in magnitude by adding an apparent distance modulus 
$(m-M)_V=13.40$. This distance modulus is consistent with the estimates $(m-M)_V=13.35\pm0.08$ \citep{thompson} 
and $(m-M)_V=13.40\pm0.07$ \citep{kalbin} from two eclipsing binaries in the cluster, and has been chosen  
to match the bottom-right end of the observed HB with models calculated 
with the normal initial He.

The increase of initial He abundance at fixed mass loss 
moves the synthetic stars towards bluer colours and brighter magnitudes. This progressive shift in colours and magnitudes 
plus the increased extension of the blue loops in the synthetic 
populations, reproduce well the wedge-shaped observed HB. For the reference age and metal composition, 
$\Delta M_{\rm RGB}$=0.23$M_{\odot}$ produces HB masses equal to 0.69, 0.66 and 0.63$M_{\odot}$ for 
$Y$=0.256, 0.270 and 0.286, respectively \citep[see also a similar discussion in][]{gratton}.
On the other hand, synthetic stars with $Y$=0.256 and increased $\Delta M_{\rm RGB}$=0.28$M_{\odot}$ (dots) are displaced towards bluer colours 
but fainter magnitudes, confirming that the morphology of the cluster HB is driven by a range of $Y$ rather than $\Delta M_{\rm RGB}$.

We have displayed also HB tracks for $Y$=0.256 and masses equal to 0.8 and 
0.9$M_{\odot}$ (in order of increasing colour and decreasing magnitude), 
that correspond to $\Delta M_{\rm RGB}$=0.12$M_{\odot}$ and 0.02$M_{\odot}$, respectively. These tracks are beyond the red edge of the observed HB, 
and no variation of the adopted distance modulus can enforce an overlap with the data.

\begin{figure}
\centering
\includegraphics[width=\columnwidth]{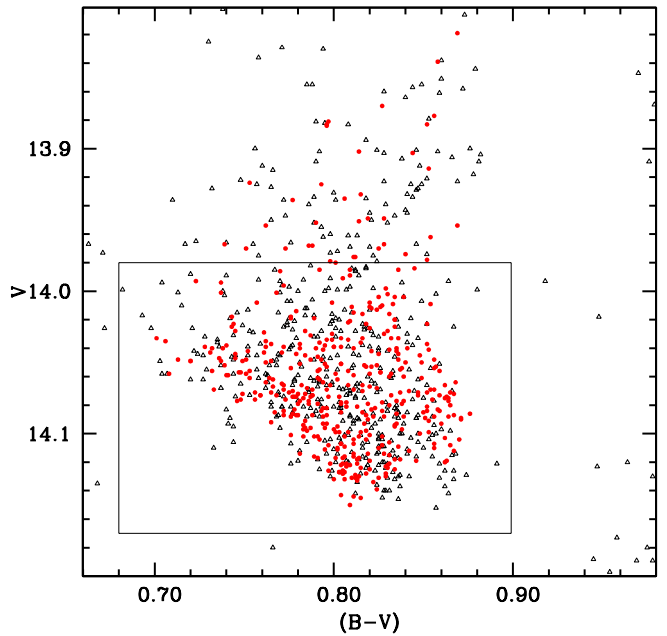}
\caption{Our baseline synthetic model for 47~Tuc HB (filled circles) compared to the observed HB (open triangles). 
The number of observed and synthetic stars within 
the box enclosing the observed HB is the same (see text for details).}
\label{HBref}
\end{figure}

A full synthetic HB compared to the observed one is shown in Fig.~\ref{HBref}. 
For this complete simulation we had to assume a 
statistical distribution for the initial He abundances, that for simplicity we considered to be uniform.
The range of $Y$ values spans the interval 0.256-0.286 ($Y_{\rm min}$=0.256 and $\Delta Y$=0.03), 
and $\Delta M_{\rm RGB}$=0.23$M_{\odot}$ like in Fig.~\ref{HBtracks}, with a 
very small Gaussian spread of 0.005$M_{\odot}$. Photometric errors have been assumed to be Gaussian, with a mean value 
equal to 0.002~mag in both $B$ and $V$ magnitudes, as obtained from the photometric data.
We restrict our comparison to the objects within the box highlighted in the CMD (the precise 
choice of the boundaries of the box is not crucial).
The simulation contains a much larger number  
of stars than observed, to minimize the Poisson error on the synthetic star counts, but here 
for the sake of clarity we show a subset of synthetic objects that matches the number of observed stars.
We considered the match to be satisfactory when the observed mean magnitude ($<V>_{\rm HB}$=14.07) and colour ($<(B-V)_{\rm HB}>$=0.80), plus 
the associated 1$\sigma$ dispersions (0.04~mag in both cases)   
are reproduced within less than 0.005~mag, 
and the overall shape of the observed star counts as a function of both $(B-V)$ and $V$
is well reproduced.  
For the assumed reddening $E(B-V)$=0.024, the mean $(B-V)$ colour is matched, and 
a distance modulus $(m-M)_V=13.40$ allows to match also the observed mean $V$ magnitude. 

\begin{figure}
\centering
\includegraphics[width=\columnwidth]{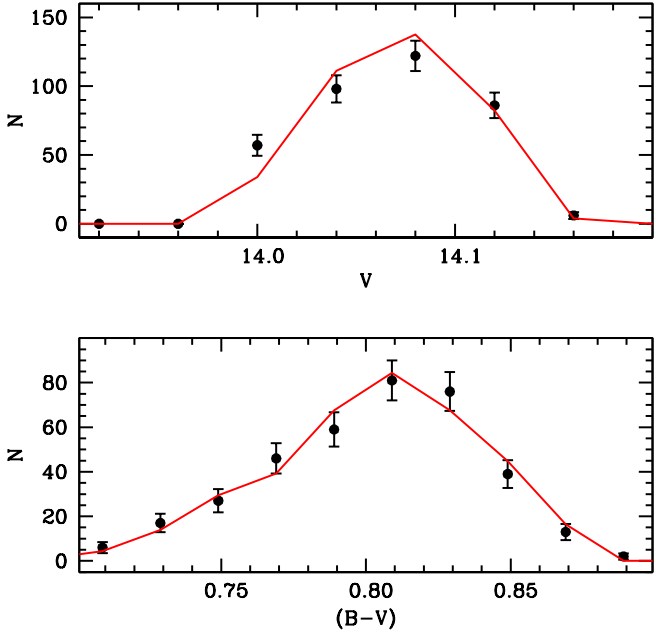}
\caption{Comparisons of synthetic (solid line) and observed (filled circles) star counts with Poisson error bars  
in $V$-magnitude (top panel -- bin size equal to 0.04~mag) and $(B-V)$ colour (bottom panel -- bin size equal to 0.02~mag) bins, 
obtained from the data in Fig.~\ref{HBref} 
(see text for details).}
\label{HBrefhist}
\end{figure}

Figure~\ref{HBrefhist} displays the resulting histograms of observed and synthetic 
star counts as a function of $V$ and $(B-V)$ (star counts from the synthetic CMD are rescaled 
to match the observed total number of HB stars). The agreement looks very good even with these simple assumptions about 
mass loss and He distribution.  

\begin{figure}
\centering
\includegraphics[width=\columnwidth]{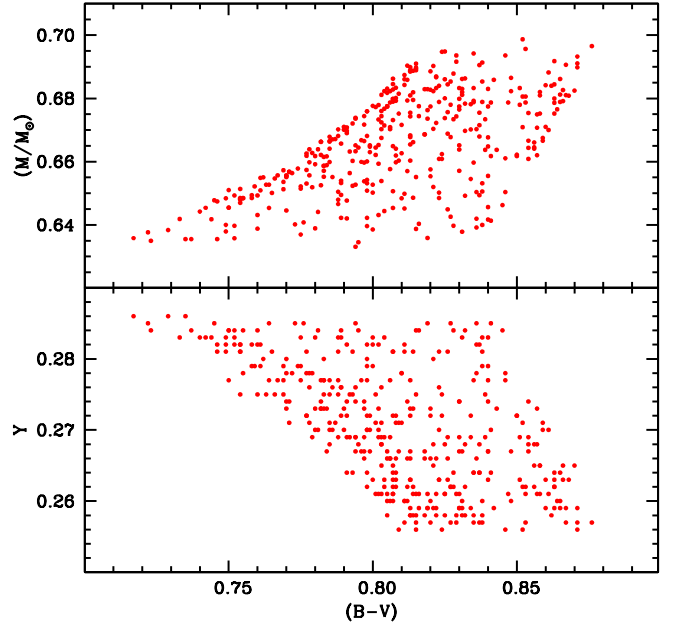}
\caption{Distribution of stellar mass (top panel) and initial $Y$ values (bottom panel) as a function of the colour  
$(B-V)$, for the stars in the synthetic CMD 
displayed in Fig.~\ref{HBref}.}
\label{HBrefdistr}
\end{figure}

Figure~\ref{HBrefdistr} displays mass and $Y$ distributions as a function of the colour of the synthetic 
stars displayed in Fig.~\ref{HBref}. There is an obvious correlation 
with $(B-V)$ for both mass and $Y$, as expected from the constant 
$\Delta M_{\rm RGB}$ (irrespective of $Y$) and very small $\sigma(\Delta M_{\rm RGB})$ assumed in the simulation, 
that is blurred by the blue loops of the HB tracks (see Fig.~\ref{HBtracks}).
This is clear from the top panel, that shows how stars with a fixed mass  
are distributed over a large range of colours.   
Redder stars on the ZAHB are on average more massive and less He-enriched than bluer objects; 
the typical mass for the stars with normal $Y$=0.256 is $\sim$0.69$M_{\odot}$, and decreases 
to $\sim$0.63$M_{\odot}$ for $Y$=0.286. The average mass along the synthetic HB is equal to 0.66$M_{\odot}$, 
with a 1$\sigma$ dispersion of 0.017$M_{\odot}$. The exact values of these two quantities depend on the assumed 
distribution of initial $Y$, but obviously the mean mass cannot be outside the range 0.63-0.69$M_{\odot}$.
A general trend of increasing $Y$ with decreasing colour is also fully consistent 
with the observed trend of increasing Na towards bluer colours along the cluster HB \citep[see][]{gratton}.

We could have tried to enforce {\sl a priori} the constraint of perfect statistical agreement between the theoretical and 
observed star counts. However, a perfect fit rests on the  
precise knowledge of the statistical distribution of $\Delta M_{\rm RGB}$ and the initial $Y$ among the cluster stars. 
Due to the current lack of firm theoretical and empirical guidance, this distribution may be extremely complicated and/or discontinuous. 
The constraints imposed on the matching synthetic HB are however sufficient to put strong constraints on $\Delta M_{\rm RGB}$ --the main parameter 
discussed in this work-- and the range of initial $Y$, which determine 
the region of the CMD covered by the observed HB. 

\begin{figure}
\centering
\includegraphics[width=\columnwidth]{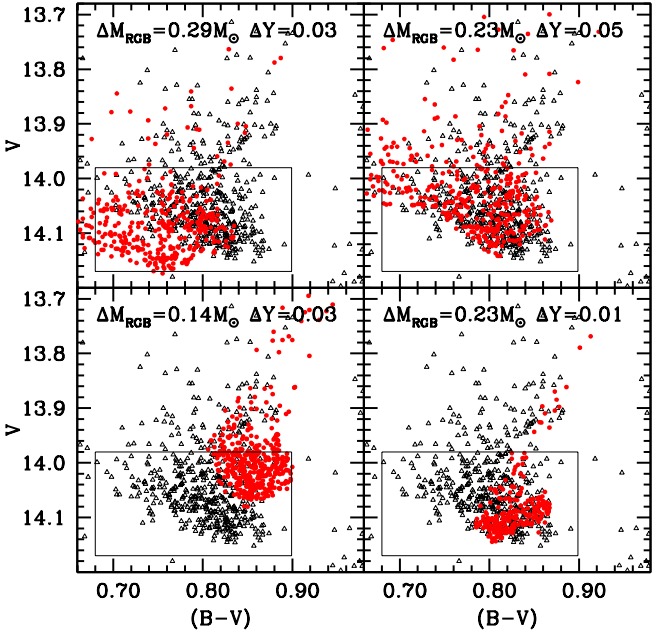}
\caption{As Fig.~\ref{HBref}, but for the labelled values of $\Delta M_{\rm RGB}$ and $\Delta Y$ (see text for details)}
\label{HBvariations}
\end{figure}

In fact, just the observed shape of the HB allows a good determination of both $\Delta M_{\rm RGB}$ and the $Y$ range, when $E(B-V)$, age 
and initial chemical composition are fixed, 
as can already be inferred from Fig.~\ref{HBtracks}. More in detail, Fig.~\ref{HBvariations} shows how changing these two parameters 
affects the shape and location of the synthetic HB. Variations of $\Delta M_{\rm RGB}$ around the reference value 
--keeping the reference $Y$ distribution fixed-- move the location of the 
synthetic HB along the direction from the top-right corner of the CMD to the bottom-left one. At the same time the HB gets compressed 
when $\Delta M_{\rm RGB}$ is reduced (shorter loops in the CMD of the HB tracks of larger mass) and stretched when $\Delta M_{\rm RGB}$ is increased.
No change of the cluster distance modulus can bring into agreement any of these two synthetic CMDs with the observed HB.
Variations of $\Delta Y$ --keeping the reference $\Delta M_{\rm RGB}$ unchanged-- stretch or compress the synthetic HB along the 
direction from the top-left corner of the CMD to the bottom-right one. Also in this case, variations of the cluster distance modulus 
do not compensate for the change of $Y$ distribution. 

A decrease of $\Delta Y$ to 0.025 and variations of $\Delta M_{\rm RGB}$ within less than 0.01~$M_{\odot}$ do still allow a satisfactory fit 
to the observed HB, according to the criteria described above, but larger variations are clearly ruled out because the resulting synthetic HB 
would clearly have a different shape than observed. Values of $\Delta M_{\rm RGB}$ below 0.20$M_{\odot}$ are totally incompatible with the 
observed HB morphology and location in the CMD.

We have also experimented with a mass loss linearly dependent on the initial $Y$ distribution. Considering $\Delta M_{\rm RGB}$=0.23$M_{\odot}$ 
for the population with $Y_{\rm min}$, a synthetic HB with $\Delta Y$=0.03 and $\Delta M_{\rm RGB}$ increasing at most as 0.5 $\times$ $\Delta Y$ 
provides a fit to the observations 
of comparable quality as the baseline simulation. 
This implies $\Delta M_{\rm RGB}$ higher by just 0.015$M_{\odot}$ for the most He-rich component.
Experiments with $\Delta M_{\rm RGB}$ decreasing with $Y$ show that at most $\Delta M_{\rm RGB}$ can decrease as 0.15 $\times$ $\Delta Y$, implying a 
negligible decrease of the RGB mass loss as a function of $Y$.

As an additional test we have calculated a synthetic HB by considering the reference $\Delta M_{\rm RGB}$=0.230$\pm$0.005 (Gaussian spread) 
for all $Y$, but with a different distribution of initial He abundances. We have considered in this case 
70\% of the stars with a Gaussian distribution of initial $Y$ characterized by mean value 
$<Y>$=0.275 and spread $\sigma(Y)$=0.007, and the remaining 30\% with a very narrow Gaussian distribution with 
$<Y>$=0.258 and spread $\sigma(Y)$=0.0008.
This choice stems from the results by \citet{milonetuc}, who found a bimodal MS for this cluster, corresponding 
to $\sim$0.02 difference in initial $Y$. The adopted distribution in our simulations has a difference of $\sim$0.02 
between the mean values of the two Gaussians, and a total range $\Delta Y$=0.03, that is needed to cover completely the $V-(B-V)$ 
region occupied by observed HB.   
The 70/30 ratio comes again from \citet{milonetuc} analysis of the number ratio between the two populations of different initial He 
as a function of the distance from the cluster centre.
For the same distance modulus of our reference simulation, mean $V$ and mean $(B-V)$ of the observed HB are again matched 
within 0.01~mag, although star counts as a function of colour and magnitude are slightly less well reproduced.

The main point of this simulation is that a change of the initial $Y$ distribution of the HB stars does not affect 
the $\Delta M_{\rm RGB}$ required to match the CMD location of the observed HB.

\section{Analysis of the uncertainties}

In the previous section we have found that for the adopted reference  
[Fe/H]=$-$0.7, [$\alpha$/Fe]=0.4, $E(B-V)$=0.024, t=11.5~Gyr, our synthetic HB simulations 
require that RGB stars lose $\Delta M_{\rm RGB}$=0.23~$M_{\odot}$ (and have a range of initial He abundances $\Delta Y$=0.03) 
to match the observed location and morphology of the HB, corresponding to stellar masses in the range 0.63-0.69$M_{\odot}$. 
This value of $\Delta M_{\rm RGB}$ (and $\Delta Y$) is broadly in line 
with previous results based on synthetic HB modeling \citep{dicri, gratton}, 
but in total disagreement with the conclusions by \citet{heyl_b} about RGB mass loss, based on cluster dynamics. 
\citet{heyl_b} results specifically exclude HB masses of the order of 0.65$M_{\odot}$.
In the following we will discuss quantitatively how our reference estimate of $\Delta M_{\rm RGB}$ may be affected by a series 
of observational and 
theoretical uncertainties.

\subsection{Cluster reddening, age and chemical composition}

The first obvious source of systematics is related to the range of reddening, chemical composition and age estimates found in the 
literature.
We analyzed these effects by varying one parameter at a time. In all these cases the value of $\Delta Y$ required to reproduce the 
observations 
is unchanged compared to the baseline case.

Increasing $E(B-V)$ (our baseline simulation has employed the lowest estimate of the cluster reddening) 
tends to increase $\Delta M_{\rm RGB}$. For example, assuming the widely employed value $E(B-V)$=0.04, we obtain 
$\Delta M_{\rm RGB}$=0.24$M_{\odot}$, 
whilst for the upper limit $E(B-V)$=0.055 we obtain $\Delta M_{\rm RGB}$=0.26$M_{\odot}$, 
e.g. we need lower masses to match the observed HB. 
As already mentioned, \citet{mmc} determined star-to-star variations around the mean reddening 
by at most $-$0.007~mag and +0.009~mag, respectively, with mean variations smaller than these extreme values.
This amount of differential reddening will hardly affect the value of $\Delta M_{\rm RGB}$, as detailed below. 

Let's assume --to maximize this effect-- 
that the stars at the red edge of the observed HB have all a reddening 0.007~mag lower than the mean value. 
To compare with theory, their colours should then be increased by 0.007~mag (and their $V$ magnitudes increased by $\sim$0.02~mag) 
to reduce the observed HB to a single value of $E(B-V)$, causing a decrease of 
$\Delta M_{\rm RGB}$ by just $\sim$0.01$M_{\odot}$ in this extreme case. 
At the same time, assuming all stars along the blue edge of the {\sl wedge} have a reddening 0.009~mag higher than the mean value, 
their colours should be decreased by the same amount and their $V$ magnitudes also decreased by $\sim$0.03~mag.
This would require a negligible increase of $\Delta Y$, and $\Delta M_{\rm RGB}$ increasing with $Y$ as 0.6 $\times$ $\Delta Y$.

As for the age, a variation by $\pm$1~Gyr around the reference value causes a change of the RGB progenitor mass 
by about $\pm$0.02 (lower mass for increasing age). The best fit synthetic HB requires 
the same HB mass distribution as the baseline case, but $\Delta M_{\rm RGB}$ varies by about $\pm$0.02 (decreased when the age increases) because of the 
change of the RGB progenitor mass.

We considered then the range of [Fe/H] estimates [Fe/H]=$-0.7\pm0.1$. 
Increasing [Fe/H] of the models tends to increase $\Delta M_{\rm RGB}$ for two reasons. First, because at fixed age 
the RGB progenitor mass  
increases by $\sim$0.01~$M_{\odot}$ for a 0.1~dex increase of [Fe/H] 
(due to longer evolutionary timescales at fixed mass) and second, HB tracks are redder, so that for a fixed reddening 
lower HB masses are required to match the observed HB location.
If we consider [Fe/H]=$-0.6$ --the approximate upper limit of the spectroscopic determinations-- keeping everything else unchanged,  
the best fit synthetic HB model has $\Delta Y$=0.03,  
$\Delta M_{\rm RGB}$=0.28$M_{\odot}$ and $(m-M)_V$=13.35. 
When employing the approximate lower limit [Fe/H]=$-0.8$, 
after interpolation in metallicity amongst the grid of BaSTI models we obtain a best match for $\Delta Y$=0.03,  
$\Delta M_{\rm RGB}$=0.18$M_{\odot}$ and $(m-M)_V$=13.47.
In these cases the mean mass of the synthetic HB stars varies by $\sim$0.03~$M_{\odot}$ (increases when [Fe/H] decreases) around the mean value of the 
baseline simulation.

If we decrease [$\alpha$/Fe] from 0.4~dex to 0.3~dex, keeping [Fe/H] and all other parameters unchanged, 
after interpolations amongst our models we found that $\Delta M_{\rm RGB}$ decreases by just 0.01-0.02$M_{\odot}$ 
(the HB mean mass changes by less than this, because of the corresponding variation of the RGB progenitor mass, as for the case of changing [Fe/H]) 
and $(m-M)_V$ increases by $\sim$0.02~mag compared to our baseline simulation.

To summarize, variations of the cluster chemical composition, age and reddening do not change much $\Delta M_{\rm RGB}$, and especially 
the typical HB stellar mass, compared to the results of the baseline simulation. In the following we discuss whether considering 
a number of uncertainties in the theoretical models can help solving or at least minimizing the disagreement with inferences from cluster dynamics.
 
\subsection{Bolometric corrections and colour transformations}

To compare the output of stellar evolution calculations ($L$ and $T_{eff}$) with observed CMDs the use of bolometric corrections (BCs) 
and colour-$T_{eff}$ relationships is essential.  
One way to proceed is to calculate grids of model atmospheres and synthetic spectra, that are   
integrated under the appropriate filter transmission functions to get fluxes in a given passband, 
which are then suitably normalized to calculate 
BCs and colours \citep[see, e.g.][and references therein]{gbb, cv}.
Our adopted set of models employs theoretical BCs and colours calculated from ATLAS9 model atmospheres and spectra \citep[see][for details]{basti, bastia}.
We have tested also the results from PHOENIX model atmosphere calculations employed by the 
Dartmouth stellar model library \citep{dotter}, and found that at the relevant metallicities the HB tracks become systematically 
redder than with ATLAS9 results. This 
shift would cause lower HB masses from synthetic HB modeling, hence higher $\Delta M_{\rm RGB}$ at fixed age, reddening and chemical composition.

An alternative approach is to employ empirical or semiempirical results, when available. 
\citet{wl} have recently presented essentially empirical BCs and colour-$T_{eff}$ relationships in various broadband filters based on 
a collection of photometry for stars with known [Fe/H]. 

\begin{figure}
\centering
\includegraphics[width=\columnwidth]{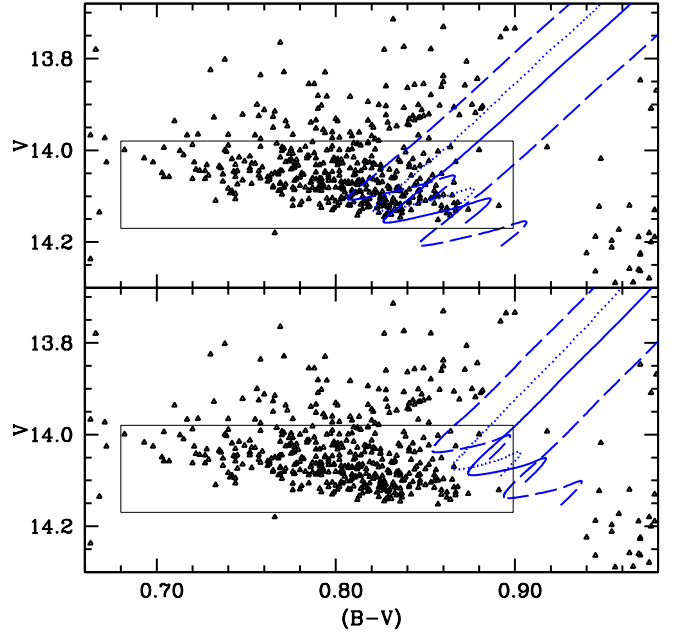}
\caption{HB evolutionary tracks for 0.7 (top panel) and 0.8$M_{\odot}$ (bottom panel), 
[Fe/H]$-$0.7, [$\alpha$/Fe]=0.4, $Y$=0.256 HB models shifted by $E(B-V)$=0.024 and $(m-M)_V$=13.40, compared 
to the cluster HB. The dotted lines display the BaSTI tracks with the adopted BaSTI BCs and colour transformations, while 
the solid lines show the same tracks but employing the empirical \citet{wl} BCs and colours. Dashed lines in each panel 
display the brightest/bluest and faintest/reddest extremes 
of the error range associated to \citet{wl} transformations.}
\label{HBbc}
\end{figure}

Figure~\ref{HBbc} displays representative 0.7$M_{\odot}$ and 0.8$M_{\odot}$ HB tracks with our adopted reference chemical composition and $Y$=0.256, 
shifted by $E(B-V)$=0.024 and $(m-M)_V$=13.40 as in our baseline simulation. 
The same tracks are displayed after applying BCs and colour transformations from \citet{wl}, using the routine 
provided by the authors. This routine gives also the errors associated to the computed BCs and colours, and 
we display in the figure the brightest/bluest and faintest/reddest limits of the area covered by these error bars, that 
are of about 0.05~mag in ${\rm BC}_V$ and $\sim$0.02~mag in $(B-V)$. 

It is clear from the figure that these transformations make the tracks redder (and very slightly fainter) 
than the reference BaSTI models, but within 
the empirical error bars ATLAS9 transformations (adopted in the BaSTI library) are consistent with \citet{wl} results.
There isn't much room for a substantial increase of the typical HB mass when considering these empirical results. 
Masses above $\sim$0.7$M_{\odot}$ along the HB are still clearly excluded. 
 
To get more quantitative results 
we have calculated synthetic HB models for the reference chemical composition and reddening 
by using \citet{wl} results.
We found a best match to the observed HB for $\Delta M_{\rm RGB}$=0.24$M_{\odot}$ --
very close to the result of our baseline simulation of Sect.~\ref{refsim}-- $(m-M)_V$=13.38, all other parameters 
being the same as in our reference simulation of Sect.~\ref{refsim}. 
By employing the bluest colours allowed by 
the error bars on \citet{wl} results --to minimize the value of $\Delta M_{\rm RGB}$
and the HB masses needed by the HB simulations-- we obtained 
$\Delta M_{\rm RGB}$=0.22$M_{\odot}$, and a mean HB mass equal to $\sim$0.67$M_{\odot}$.
Considering also the errors on ${\rm BC}_V$ would simply shift 
the synthetic HB vertically by $\pm$0.05~mag, implying an adjustment of the derived $(m-M)_V$ by the same amount, that is 
still within the errors of the eclipsing binary distance estimates. 


\subsection{Model calculation, input physics}

Before the start of the HB phase GC stars go  
through the violent core helium flash at the tip of the RGB. Until recently only few calculations have been able to calculate the evolution of stellar models 
through this event, and traditionally HB models (including the BaSTI models employed in our analysis) have been 
computed by starting new sequences on the HB, where the initial structure is taken from that at the tip of the RGB. 
The underlying assumption of such methods (consistent with the results of stellar evolution calculations that follow 
the helium flash evolution) is that during the helium flash the internal (and surface) chemical structure is not altered significantly, apart from a 
small percentage 
of C produced in the He-core during the flash. 

The technique employed in BaSTI models \citep[denoted as Method 2 in the work by][]{sw05} envisages that 
a model at the beginning of the core helium flash is employed as the starting model for the ZAHB calculation. 
The core mass and chemical profile of the initial pre-flash configuration are kept unchanged. 
The total mass of the ZAHB model 
is either preserved or reduced (to produce a set of ZAHB models of varying total mass) by rescaling the envelope mass.
The new model on the ZAHB is then converged and {\sl relaxed} for a certain amount of time (typically 1~Myr) to attain CNO equilibrium in 
the H-burning shell \citep[see, e.g.,][]{cs13} before being identified as the new ZAHB model.
A 5\% mass fraction of carbon is added to the He-core composition, guided by requirement that the energy needed for the expansion of 
the degenerate helium core must come from helium burning \citep[][]{ir70}. 

\citet{vdb} and with more details \citet{piersanti} and \citet{sw05}, have compared the ZAHB location and HB evolution of models 
whose RGB progenitor evolution was properly followed through the He-flash, with results with our method described before. 
Especially for red HB models, like the case of 47~Tuc HB, differences in $T_{eff}$, $L$ and time evolution turned out to be negligible.

\begin{figure}
\centering
\includegraphics[width=\columnwidth]{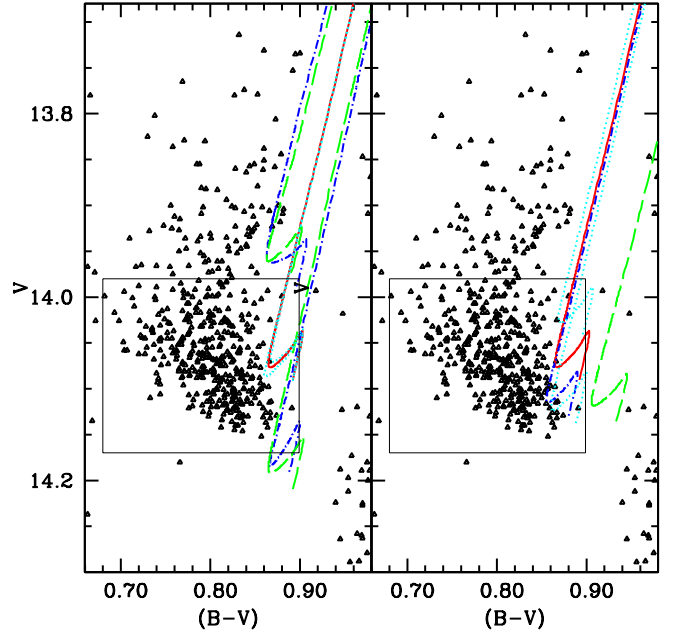}
\caption{As Fig.~\ref{HBbc} but displaying for the 0.8$M_{\odot}$ standard BaSTI track (solid line in both panels), the effect of changing  
the He-core mass at the He-flash by $\pm$0.02~$M_{\odot}$ (dashed lines in the left panel), the amount of He dredged to 
the surface by the first dredge up 
by $\delta Y_{FDU}=\pm0.02$ (dash-dotted lines in the left panel), the ${\rm ^{12}C+\alpha}$ reaction rate (doubled and halved, respectively -- 
dotted lines in the left panel),  
the mixing length $\alpha_{\rm MLT}$ by $-$0.15 (dashed line in the 
right panel), the Rosseland mean opacities by $\pm$5 \% (dotted lines in the right panel) and the ${\rm ^{14}N+p}$ reaction rate from the \cite{nacre} 
to the \cite{luna} tabulations (dash-dotted line in the right panel). }
\label{HBmcoreextHe}
\end{figure}

As additional potential sources of uncertainty in the determination of $\Delta M_{\rm RGB}$ employing synthetic HB modeling, we considered the effect 
of uncertainties in the current input physics adopted in stellar model calculations. We considered first the effect on the He-core mass at the He 
flash and how 
this affects the HB tracks, and then the effect on the HB evolution at fixed core mass.   

Since the calculation of the BaSTI model database, two relevant physics inputs, i.e. the ${\rm ^{14}N+p}$  reaction rate and the 
electron conduction opacities, have been 
subject to revised and improved determinations. The recent \cite{cpp} calculations of electron conduction opacities 
--larger than the \cite{potek} results used in the BaSTI calculations-- 
cause a variation of the He-core mass at the He flash 
($\Delta M_{He_c}$) by $-$0.006$M_{\odot}$, compared to our adopted HB models. On the other hand, the new \cite{luna} tabulations of the ${\rm ^{14}N+p}$ reaction rate 
--lower by a factor $\sim2$ than the \cite{nacre} rate used in our adopted models-- 
induce an increase $\Delta M_{He_c} \sim$0.0025$M_{\odot}$ compared to the BaSTI calculations. The combined effect is a minor change 
$\Delta M_{He_c}\sim-$0.0035$M_{\odot}$ compared to the BaSTI models.

In addition, \cite{vopd} have determined $\Delta M_{He_c}$ due to realistic uncertainties in the 
\cite{luna} ${\rm ^{14}N+p}$ reaction rate ($\pm$10\%), the \cite{cpp} electron conduction opacities ($\pm$5\%), 
plus the ${\rm ^{1}H+p}$ ($\pm$3\%) and 3$\alpha$ ($\pm$20\%) reaction rates, neutrino emission rates ($\pm$4\%)
and the radiative opacities ($\pm$15\%) used in the BaSTI models. If we simply add together the effect of these uncertainties plus  
the systematic effect of including fully efficient atomic diffusion in the pre-HB calculations\footnote{The BaSTI models neglect atomic diffusion} 
--that increases $M_{He_c}$ by $\sim0.004M_{\odot}$, see Table~7.2 in \cite{cs13}-- and the $\Delta M_{He_c}\sim-$0.0035$M_{\odot}$ discussed before, 
we obtain a (conservative) variation $\Delta M_{He_c}\sim \pm$0.01$M_{\odot}$ around the value obtained from the BaSTI calculations.

The left-hand panel of Fig.~\ref{HBmcoreextHe} displays the effect of $\Delta M_{He_c}$=$\pm$0.02$M_{\odot}$ 
on a 0.8$M_{\odot}$ HB track for the reference chemical composition $Y$=0.256, [Fe/H]=$-$0.7 and [$\alpha$/Fe]=0.4. 
To calculate this track we varied the He-core mass 
by $\pm$0.02$M_{\odot}$ around the value provided by the BaSTI models ($M_{He_c}=0.48 M_{\odot}$) and recalculated the HB evolution, 
everything else being kept fixed.
This variation is even larger than the estimates discussed before and takes into account two additional factors. 
First, a possible 
extra increase of $M_{He_c}$ due to the effect of rotation, as originally discussed by \cite{ct} and \cite{ldz} in the context of synthetic HB modeling; second, 
a potential additional decrease by $\sim$0.008$M_{\odot}$ due to the use of weak screening in the He-core during the whole RGB evolution, instead of the 
transition to intermediate and strong screening when appropriate, following the treatment by \cite{grabo} as implemented in the BaSTI calculations. 

It is clear from the figure that 
the expected shift in magnitude of the tracks (higher $M_{He_c}$ corresponds to higher luminosity) is not accompanied by 
any major shift in colour. The ZAHB location is only very slightly redder for the lower core mass model, and bluer for the higher core mass one, and the extension 
of the loops in the CMD is only slightly altered. 
It may seem puzzling that such a large variation of $M_{He_c}$ hardly affects the tracks. In fact HB tracks 
of a given total mass 
are generally sensitive to small variations of the He-core mass at the flash, because 
of the changed efficiency of the H-burning shell, due to the corresponding variation of the mass of the H-rich envelope.
However, at this metallicity and for massive HB models, the efficiency of the burning shell 
is only weakly altered even by a $\pm$0.02$M_{\odot}$ change of $M_{He_c}$, because this corresponds to just a small 
percentage variation of the mass thickness of the envelope.

The same figure shows also the effect of arbitrarily altering the efficiency of the first dredge-up on the same 0.8$M_{\odot}$ HB track, 
keeping everything else unchanged. 
Our adopted BaSTI models predict an increase of the surface He mass fraction by 0.02 for the cluster RGB stars. 
This increased He abundance in the envelope impacts the efficiency of the H-burning shell during the HB phase. 
We tested a variation of the dredged up He mass fraction  
$\delta Y_{FDU}=\pm0.02$, by recalculating HB models with this new envelope chemical abundance; the resulting tracks  
are simply shifted in luminosity (higher for increasing He abundance) but their colour location is unchanged. 

We have then considered the HB evolution of the same 0.8$M_{\odot}$ HB track keeping 
$M_{He_c}$ fixed, and varying once at a time the most relevant inputs like 
the radiative opacities by $\pm 5\%$, the ${\rm ^{14}N+p}$ reaction rate from the \cite{nacre} 
tabulations used in our adopted BaSTI models to the most updated \cite{luna} tabulations, the 
superadiabatic convection mixing length parameter by $\delta \alpha_{\rm MLT}$=$-$0.15 
compared to the solar calibrated value ($\alpha_{ml, \odot}=2.01$) of the BaSTI calculations (see right-hand panel of Fig.~\ref{HBmcoreextHe}) 
and the ${\rm ^{12}C+\alpha}$ reaction rate (doubled and halved, respectively, as shown in the left-hand panel of Fig.~\ref{HBmcoreextHe}).

As already mentioned, the \cite{luna} ${\rm ^{14}N+p}$ rate is about a factor of two lower than the \cite{nacre} result, i.e., a variation much larger 
than the error associated to this improved rate. Regarding the variation $\delta \alpha_{\rm MLT}$=$-$0.15 with respect to the solar calibrated value, 
this is what the fitting formulas by \cite{mwa} --based on the results of a large grid of 3D radiation hydrodynamics simulations-- predict for the 
surface gravity-effective temperature regime of the 0.8$M_{\odot}$ HB track, and a metallicity [Fe/H]=$-$0.5, close to the value adopted in 
our calculations.
We consider this adopted $\delta \alpha_{\rm MLT}$ as a qualitative estimate of the uncertainty on the efficiency of superadiabatic convection 
in the envelope of red HB stars. Regardless of the precise 
estimate of $\delta \alpha_{\rm MLT}$, \cite{mwa} simulations \citep[a similar behaviour is found also in an 
analogous 3D hydro-calibration by][at solar metallicity]{tr} predict that  
$\alpha_{\rm MLT}$ should decrease towards higher effective temperature and/or lower surface gravity compared to the solar values.
As for the ${\rm ^{12}C+\alpha}$ reaction rate, we consider both an increase and a decrease by a factor of two with respect to the reference 
rate adopted in the BaSTI calculations \citep[][]{kunz}, along the lines of the analysis by \citet{gai} regarding the current uncertainties 
on this reaction rate.

It is clear from the figure that the variation of the opacity (higher opacity corresponds to lower luminosities) and ${\rm ^{14}N+p}$ reaction rate 
alter essentially just the brightness of the model, not the colour, as for the case of varying  $M_{He_c}$ and $Y_{FDU}$. 
The variation of the ${\rm ^{12}C+\alpha}$ reaction rate alters only very marginally the extension of the blueward 
loop in the CMD (more extended loop for an increase of the reaction rate).
The variation of the mixing length obviously affects the model colours, but the decrease of $\alpha_{\rm MLT}$ predicted by the hydro-simulations 
makes the model redder, thus decreasing the mass of the HB models --hence increasing the 
RGB mass loss-- needed to match the observed HB.
\footnote{Somewhat surprisingly also the $V$-band magnitude of the model with changed mixing length is affected, although this is due exclusively to 
the bolometric corrections, that change because of the decrease of the model effective temperature}. 

Finally, we considered the effect of the treatment of core mixing during the HB phase. As well known and 
discussed recently again by, e.g., \cite{gnm} and \cite{consta}, 
the treatment of convective boundaries during core helium burning is still an open question in stellar evolution calculations, and is handled 
in different ways by different stellar evolution codes. Recent advances in asteroseismic observations and techniques 
\citep[][]{bossini, consta} are starting to add very direct observational constraints to the core mixing process during the central He-burning phase, that 
coupled to theoretical inferences and indications from star counts in Galactic globular clusters \citep[see, e.g.,][and references therein]{cct, gnm, csi}, 
make a strong case for the core mixed region to be extended beyond the Schwarzschild border.
Questions exist however about the treatment of this extended mixing.
The details do not affect the ZAHB location of the models, but can potentially modify the extension of the 
loops in the CMD and have some relevance for the determination of the mass range of the cluster HB population.

Our adopted BaSTI HB models include semiconvection in the core \citep[see][]{cgra,cgrb, cctp}, with the suppression of the breathing pulses 
in the last phases of central He-burning with the technique by \cite{cct}, following the observational constraints discussed by \cite{csi}. 
\cite{constantino} have calculated and compared HB models with different treatment of core mixing beyond the Schwarzschild border, and 
in their Fig.~7 they compare HB tracks for a 0.83$M_{\odot}$, [Fe/H]=$-$1, initial $Y$=0.245, not very different from the case discussed 
in this section. 
Tracks with two different types of overshooting \citep[one of them, the {\sl maximal overshoot} case,  
reproduces best the asteroseismic constraints, as discussed by][]{consta} and with semiconvection are almost indistinguishable. 

As a conclusion, none of the large list of uncertainties discussed in this section, 
appears to be able to shift onto the observed HB substantially more massive HB tracks, compared 
to the baseline simulation.


\section{Discussion and conclusions}

In the previous sections we have investigated in detail the constraints posed by the CMD of HB stars on the RGB mass loss 
in the Galactic globular cluster 47~Tuc, using synthetic HB modeling.
We confirm the results by \citet{dicri} and \citet{gratton} about the need for a range of initial He abundances (we find $\Delta Y$=0.03) 
to reproduce properly the 
observed HB morphology. This abundance range is broadly consistent with the range inferred from the analysis of the cluster MS.  
For the values of age (11.5~Gyr), metal content ([Fe/H]=$-$0.7 and [$\alpha$/Fe]=0.4)  
and reddening ($E(B-V)$=0.024) adopted in our baseline simulation, 
a total mass loss of individual RGB stars $\Delta M_{\rm RGB}$=0.23$M_{\odot}$ (with a very small 
dispersion) independent of $Y$ is required to reproduce the observed HB location and morphology.
This value for $\Delta M_{\rm RGB}$ is consistent with the estimates by \citet{ori07} from mid-IR photometry and the results 
by \citet{dicri} and \citet{gratton} from similar synthetic HB modeling. The mean mass of the synthetic HB stars 
in this simulation is equal to 0.66$M_{\odot}$.

To assess the robustness of this estimate of $\Delta M_{\rm RGB}$ and the HB masses,  
we have then considered several possible sources of uncertainty that could affect the HB modeling, stemming from uncertainties in 
both cluster parameters 
and model calculations. 
Uncertainties in the cluster reddening, age and [Fe/H] have the 
largest effect on the derived $\Delta M_{\rm RGB}$. Considering and age of 12.5~Gyr (approximately the 
upper limit consistent with current estimates), [Fe/H]=$-$0.8 (the lower limit of spectroscopic estimates) 
and the lowest reddening estimate $E(B-V)$=0.024 could potentially lead to $\Delta M_{\rm RGB} <$0.20$M_{\odot}$. 
Younger ages, higher metallicities and reddenings, within the range of current estimates, can 
however increase the estimated  $\Delta M_{\rm RGB}$ up to $\sim$0.30$M_{\odot}$. A fraction of this range of $\Delta M_{\rm RGB}$ values 
is due to the variation of the RGB progenitor mass with age and initial chemical composition, so that the actual variation of 
the typical HB masses is smaller than the full range of $\Delta M_{\rm RGB}$.

\begin{figure}
\centering
\includegraphics[width=\columnwidth]{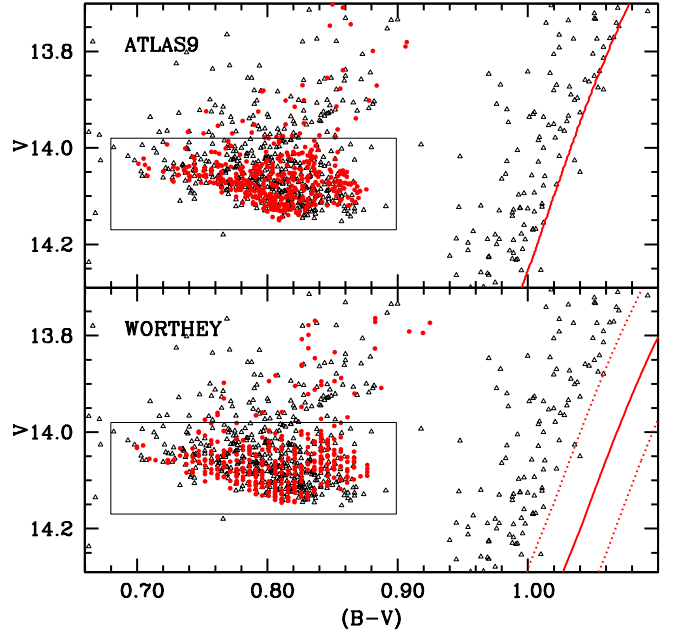}
\caption{As Fig.~\ref{HBref} but including also the observed and theoretical RGB sequences. 
The top panel displays the [Fe/H]=$-$0.7 baseline simulation 
employing the BaSTI adopted BCs and colour transformations (ATLAS9). The bottom panel shows 
the result for the baseline  
simulation but employing the empirical \citet{wl} BCs and colour transformations (solid line). The reddest and bluest 
limits of the RGB colours according to the errors on the \citet{wl} transformations and BCs are displayed 
as dotted lines (see text for details).
}
\label{HBRGB_a}
\end{figure}

There is however the additional constraint posed by the observed colour of the RGB that can be considered. 
As we show in the following, this 
will narrow the range of $\Delta M_{\rm RGB}$ and HB masses allowed by the analysis of the cluster CMD.
The upper panel of Fig.~\ref{HBRGB_a} shows the synthetic CMDs of our baseline simulation described in Sect.~\ref{refsim} 
([Fe/H]=$-$0.7, [$\alpha$/Fe]=0.4, $E(B-V)$=0.024, age t=11.5~Gyr, $\Delta Y$=0.03, $\Delta M_{\rm RGB}$=0.23$M_{\odot}$, $(m-M)_V=13.40$) together 
with the corresponding RGB isochrone (also from the BaSTI models, using the reference ATLAS9 transformations), 
compared to the cluster photometry, in the magnitude range around the HB.
Obviously the theoretical RGB is too red compared to the data; given that $E(B-V)$=0.024 is the lowest reddening estimate, 
there is no room for shifting to the blue the position of the theoretical RGB (that is virtually insensitive to age and initial $Y$  
for the relevant age and He-abundance ranges).   
This implies that our adopted models with [Fe/H]=$-$0.7 are inconsistent with the observations.
 
The lower panel of Fig.~\ref{HBRGB_a} displays the best match synthetic HB simulation when applying 
the \citet{wl} transformations to our adopted BaSTI stellar models, keeping [Fe/H]=$-$0.7, [$\alpha$/Fe]=0.4, 
$E(B-V)$=0.024, t=11.5~Gyr, 
as in the baseline simulation of the top panel. We derive in this case, as already discussed,$\Delta Y$=0.03,   
$\Delta M_{\rm RGB}$=0.24$M_{\odot}$, $(m-M)_V=13.38$. The position of the RGB is again largely inconsistent 
with observations, even allowing for the errors on the \citet{wl} BCs and colour transformations, displayed in the figure.

\begin{figure}
\centering
\includegraphics[width=\columnwidth]{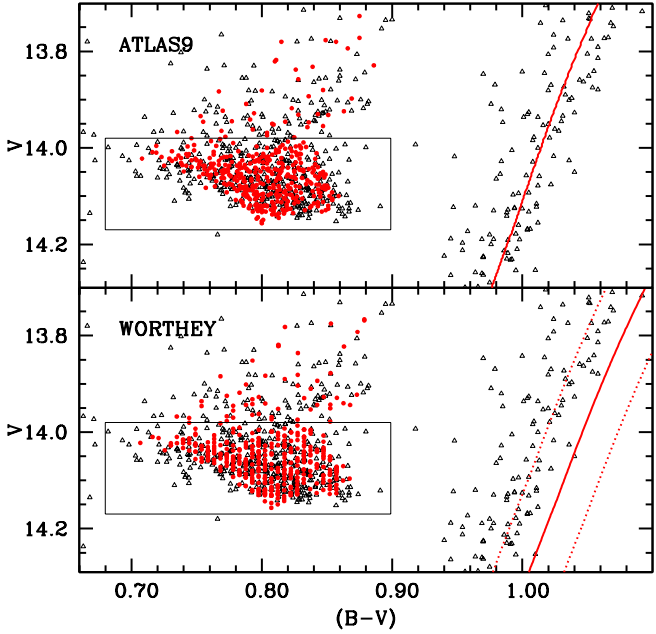}
\caption{As Fig.~\ref{HBRGB_a}, but for simulations with [Fe/H]=$-$0.8 (see text for details).}
\label{HBRGB_b}
\end{figure}

Figure~\ref{HBRGB_b} shows an analogous comparison, but for [Fe/H]=$-$0.8 ([$\alpha$/Fe]=0.4), t=11.5~Gyr. 
In the top panel --the case with the ATLAS9 BCs and colours-- the theoretical RGB matches the average colour of observed one for $E(B-V)$=0.035, 
and the best match synthetic HB has $\Delta Y$=0.03, $\Delta M_{\rm RGB}$=0.19$M_{\odot}$, $(m-M)_V=13.46$.
The average mass along the synthetic HB is in this case equal to 0.69$M_{\odot}$, 
with a 1$\sigma$ dispersion of 0.017$M_{\odot}$, and a full range between $\sim$0.65$M_{\odot}$ and $\sim$0.73$M_{\odot}$.  

The bottom panel displays the case with the \citet{wl} transformations, 
for [Fe/H]=$-$0.8 ([$\alpha$/Fe]=0.4), t=11.5~Gyr, $E(B-V)$=0.035.   
The only way to match the observed RGB is to consider 
the bluest colors allowed by the errors on these empirical transformations, whilst for the HB the reference BCs and colour transformations  
provided by \citet{wl} allow a match with observations for   
$\Delta Y$=0.03, $\Delta M_{\rm RGB}$=0.19$M_{\odot}$, $(m-M)_V=13.44$, almost identical to the case with the ATLAS9 transformations.

One can speculate whether considering the bluest limit of \citet{wl} colour transformations also for the HB models, might decrease 
$\Delta M_{\rm RGB}$, because the synthetic HB should then made redder while keeping reddening, age and chemical composition fixed. 
However, an increase of the HB masses --hence a decrease of $\Delta M_{\rm RGB}$ below $\sim$0.19$M_{\odot}$-- would cause 
a shift to the red of the HB, but also a change of 
the HB morphology that is inconsistent with the observations, like the case discussed in the lower-left panel of Fig.~\ref{HBvariations}. 
With increasing HB mass the loops in the CMD do shrink, and it is impossible to match at the same time 
the vertical and horizontal thickness of the observed HB by changing the $Y$ distribution.

In summary, the additional constraint posed by the colour of the RGB narrows down the possible range 
of [Fe/H], $E(B-V)$ and $\Delta M_{\rm RGB}$. 
A fit for the minimum estimate $E(B-V)$=0.024 could be in principle achieved, at least 
with the ATLAS9 transformations, considering a slightly higher [Fe/H], between $\sim -$0.8 and $\sim -$0.75~dex, 
but this would keep $\Delta M_{\rm RGB}$ (and the evolving HB masses) roughly unchanged because of the compensating effects of increasing [Fe/H] 
and decreasing 
reddening, discussed in the previous sections.

A firm estimate of the minimum RGB mass loss allowed by our modeling is therefore 
$\Delta M_{\rm RGB}\sim$0.17$M_{\odot}$ 
if we consider a cluster age of 12.5~Gyr, that is approximately the upper limit of current estimates. 
The maximum $\Delta M_{\rm RGB}$ is equal to $\sim$0.21$M_{\odot}$, 
for an age of 10.5~Gyr. The resulting mass distribution of the HB stars is in both cases between 
0.65$M_{\odot}$ and 0.73$M_{\odot}$, the full range determined mainly by the range of initial $Y$ abundances of 
the progenitors ($\Delta Y$=0.03). The derived  
distance modulus is consistent with constraints from the observed cluster eclipsing binaries.

We can also compare $\Delta M_{\rm RGB}$  with the expected mass loss during the following AGB phase ($\Delta M_{\rm AGB}$).
In the simulation that minimizes $\Delta M_{\rm RGB}$, the initial progenitor mass of the HB stars is equal to 
$\sim$0.89$M_{\odot}$ for the $Y$=0.254 population, and $\sim$0.84$M_{\odot}$ for the $Y$=0.284 population. 
Assuming that the final WD mass will be equal to $\sim$0.53$M_{\odot}$ \citep{kalwd} irrespective of 
the initial $Y$, the mass to be shed 
by the cluster stars during the AGB phase will range between $\Delta M_{\rm AGB}\sim$0.19$M_{\odot}$ for the $Y$=0.254 population, 
and $\sim$0.14$M_{\odot}$ for the $Y$=0.284 population. 
The value of $\Delta M_{\rm RGB}$ is therefore expected to be similar to the average $\Delta M_{\rm AGB}$, whose precise value 
depends on the distribution of initial $Y$. For a flat $Y$ distribution the average $\Delta M_{\rm AGB}$ is equal to $\Delta M_{\rm RGB}$. 

On the whole our detailed analysis confirms the discrepancy between information coming from cluster dynamics and CMD modeling of the HB.
The lower limit for $\Delta M_{\rm RGB}$ allowed by the HB modeling is only slightly lower than the 0.20$M_{\odot}$ value 
excluded with high confidence by \citet{heyl_b} analysis. The predicted mass distribution of HB stars is between 
$\sim$0.65$M_{\odot}$ and $\sim$0.73$M_{\odot}$, with mean value --that depend on the exact $Y$ distribution-- 
not much higher than 0.65$M_{\odot}$, this latter excluded by \citet{heyl_b} analysis.
On the other hand the RGB mass loss allowed by the HB modeling is consistent with  
$\Delta M_{\rm RGB}\sim 0.23\pm0.07 M_{\odot}$ estimated from the detection of their circumstellar envelopes by means of mid-IR photometry 
\citep[][]{ori07}. 

A comparison between the results from these two techniques applied to other clusters is required, 
to gain more insights about the origin of this apparently major disagreement between CMD and dynamical results.


\begin{acknowledgements}
We thank the referee for comments that pushed us to improve the analysis of Section 3.3. 
SC warmly thanks the financial support by PRIN-INAF2014 (PI: S. Cassisi), and the Economy and Competitiveness Ministry of the 
Kingdom of Spain (grant AYA2013-42781-P).
A.P. acknowledges  financial support by PRIN-INAF 2012 {\sl The M4 Core Project with Hubble Space Telescope} (P.I.: L. Bedin) 
\end{acknowledgements}

\bibliographystyle{aa}

\end{document}